# Experimental study of the flow structure stability on the bubble surface

**Anastasia Shmyrova and Andrey Shmyrov**

Institute of Continuous Media Mechanics, Ak. Koroleva St. 1, 614000, Perm, Russia

E-mail: lutsik@icmm.ru

**Abstract**. The results of the flow structure visualization experiments conducted on the surface of a single bubble streamlined by uniform flow are presented. It is shown that, at certain critical values for bubble size, flow velocity, and contamination level, the axial symmetry of the surface flow loses its stability in a threshold manner, and the first instability mode in the form of two vortices appears. Below the threshold, the stationary flow on the bubble surface is impossible. The experimental results indicate that the assumption about the axial symmetry of the motion on the bubble surface containing surfactants, which is used in most theoretical and numerical studies, is invalid. Analysis of the results has revealed the most likely reason for the spiral form of the trajectory in the problem of a small rising bubble in the surrounding fluid. For the surfactant-free surface realized in the experiments with isopropyl alcohol, the rising trajectory was a straight line, and no vortex structures were observed on the bubble surface. In the experiments with water, a spiral rising trajectory was observed, and the first instability mode was formed on the bubble surface.

## 1. Introduction

Multiphase media are the subject of intense research, with diverse applications ranging from mining, chemical, oil and gas processing to food industry. They are also widely used in bio - and pharmaceutical industries, in extraction and rectification, bubbling and flotation, intensification of chemical reactions, and production of porous and composite materials. The study of multiphase media is at the junction of physical chemistry with hydromechanics, and this generates considerable difficulties. On the one side, the presence of an interfacial surface requires information on adsorption and desorption and, consequently, a deep knowledge of the kinetics of these processes on the surface and other physicochemical properties of the liquids and gases involved in these processes. On the other hand, it is necessary to have a comprehensive insight in mechanics and thermodynamics of multiphase media. The behavior of systems with surfactants requires special attention. The existence of mutual feedback between the fields of liquid velocity and surfactant concentration at the interface makes it difficult to take into account the complexity of capillary effects. Analytical solutions are currently available for a small number of simple model approximations [1, 2].

The motion of a bubble or a drop of one liquid in another is one of the most studied problems in the nonequilibrium mechanics of multiphase media containing surfactants. The problem, though seemingly simple, still does not have an analytical solution. Because of the high resource consumption, numerical modeling is performed with discrete parameters under the assumptions about the axial symmetry of the flow on the surface of the rising bubble [3–5]. For this reason, the problem cannot be comprehensively analyzed for stability both by methods of perturbation theory and direct numerical modeling.

In the studies on thermal and concentration-capillary Marangoni convection from a local source in the presence of the adsorbed surfactant layer [6–15], it was shown that, regardless of the nature of forces that cause the liquid to move, there occurs transformation of a radial flow to a vortex flow. In our opinion, the vortex motion should also be observed in the problem of rising bubbles and droplets in the presence of surfactants. Indirect evidence of this is the velocity of the rising bubbles which was measured in a number of experimental studies [16–22]. It turns out that this velocity is smaller than that predicted by the Hadamard–Rybczynski theory and greater than the value found from the Stokes equation.

In this paper, we describe the results of the flow structure visualization experiments conducted on the surface of a single bubble in an axially symmetric flow. It has been found that, at certain critical values for bubble size, flow velocity, and contamination level, the axial symmetry of the surface flow loses its stability in a threshold manner, and the first instability mode in the form of two vortices appears. The latter refutes the assumptions about the axial symmetry of the flow made in the mathematical models known to the authors. These experimental data can be used to verify the theory.

## 2. Experimental setup

An experimental study to investigate the conditions leading to the development of three-dimensional instability on the surface of the streamlined by axially symmetrical flow spherical bubble was carried using a motionless gas inclusion in the laboratory coordinate system. The volume flow was organized in two independent ways (figure 1). In the first case (figure 1(a)), a bubble *1* with a diameter of 1-2 mm was created at the end of a thin capillary *2* (diameter 0.7 mm, length 60 cm). A capillary was fixed on the axis of a glass cylinder *3* (diameter 4 cm, length 60 cm). A vessel filled with high purified water was located on the platform of a movable support *4*. Using a stepper motor, the platform was driven along a vertical guide *5*. The periodic rectangular signal generated by an acoustic generator determined the movement speed. The uniform movement of the cylindrical vessel as a whole made it possible to create an axially symmetrical flow relative to the bubble along the entire cross-section of the channel. To this end, the capillary was positioned parallel to the stream lines with a deviation of less than 30 arc minutes. The experimental setup involved a very limited number of elements coming into contact with water, which made it easier to keep a clean experimental setup. This flow organization has the following disadvantages: the time of the experimental study is bounded by the length of the vessel, and as a result a narrow range of flow rates can be realized.

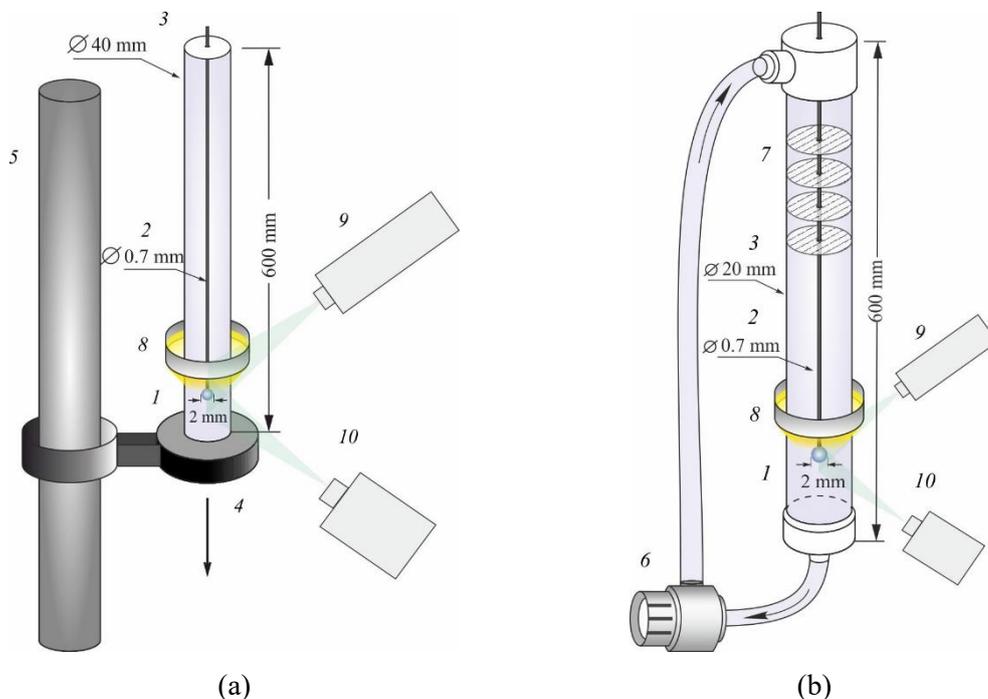

(a) (b)

**Figure 1.** Scheme of the dynamic cylinder (a) and closed contour (b) Numbers indicate parts of the setup: *1* – bubble, *2* – capillary, *3* – glass cylinder, *4* – carriage, *5* – vertical guide plate, *6* – centrifugal chemical pump, *7* – system of brass grids, *8* – ring lighting, *9* – laser, *10* – camera.

In the second method of the flow organization (figure 1(b)), a bubble *1* was created at the end of the capillary *2* placed on the axis of the glass cylinder *3* (diameter 2 cm, length 60 cm). The flow was generated by pumping high purified water in a closed loop using a centrifugal pump *6*. The system of brass grids *7* with a hole size of 0.5×0.5 mm was used for for flow laminarization. It was positioned in the upper part of the channel. This system allows conducting long-term experiments with constant control parameters within a wider range of flow rates. The flow organized in this system has a Poiseuille velocity profile because the fluid motion is realized due to a pressure drop, which requires more strict orientation of the capillary relative to the axis of the cylinder. The disadvantage of this setup is the time-consuming cleaning procedure, which, as will be shown below, does not allow us to implement the case of a clean water-air interface. Before assembling and filling with the working fluid, both systems were cleaned with a chrome mixture and rinsed with the high purity water.

Visualization of the flow structure on the bubble surface was made with polyamide particles tracer deposited at the interface in the scattered ring-shaped illuminating light 8 (figure 1). The near-surface flow was visualized by dye-streaks method and the sheet of laser light from laser module 9. The sheet plane was parallel to the velocity vector of the main flow. Video recording was carried out using a video camera 10. The use of the methods described above made it possible to detect vortex structures and to determine an azimuthal wave number. During the experiments, the change in the bubble size was compensated by gas injection (or outgassing) into the air channel with the aid of a syringe pump. Water and isopropanol were used as working fluids.

In order to realize the classical dye-streak method, several holes *2* with a diameter of 0.1 mm were made along the perimeter of the capillary *1* at a height of about 3 mm (figure 2(a)). Bubbles *3* were generated by supplying air through a fluoroplast tube *4* inside the capillary *1*. In the experiment, an aqueous solution of an inorganic dye (chromium(II) chloride) was fed into the gap between the inner walls of the capillary *1* and the tube *4* using an additional syringe pump. This dye does not exhibit surface-active properties and, when added to water, does not contaminate the interfacial surface. The color video camera was used to record the resulting jets *5*.

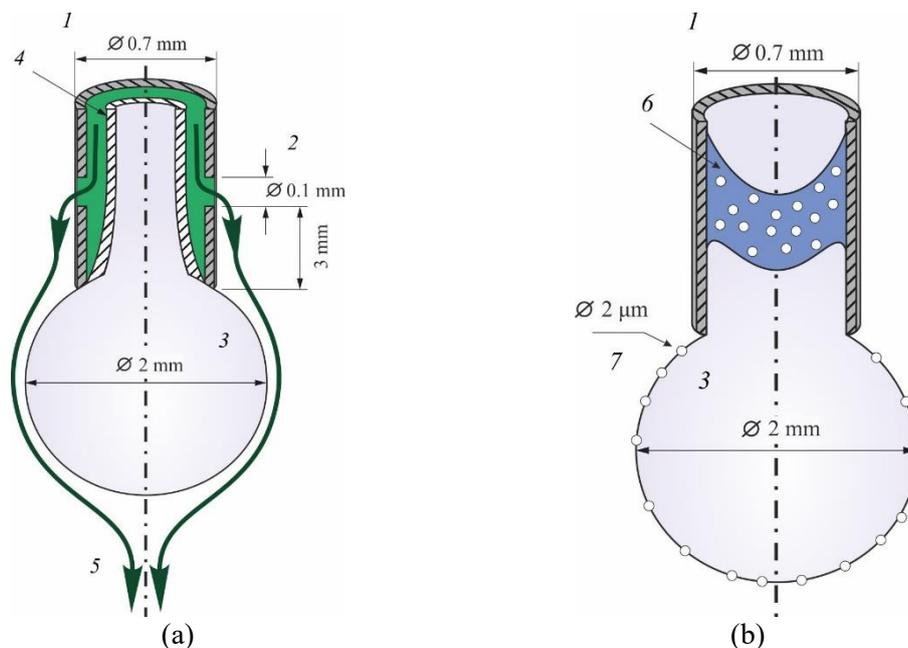

**Figure 2.** An example of application of flow visualization using' the dye-streak (a) and particle tracer methods (b): *1* – capillary, *2* – hole, *3* – gas bubble, *4* – fluoroplast tube, *5* – dye streaks, *6* – liquid bridge, *7* – polyamide particle tracer.

The second visualization method used in this study includes adding light-scattering tracers (figure 2(b)). Direct injection of tracers into the liquid volume does not allow us to visualize the motion at the interface, since the hydrophobic particles do not fall on the surface of the formed bubble. Air and an aqueous solution of particles were fed into the capillary *1* using a two-channel pump through two independent syringes. Immediately before entering the capillary, these flows mixed to form periodically successive air bubbles separated by liquid bridges of water with particles *6*. When creating a bubble, the water film was pushed out of the capillary, thereby evenly filling the interface with particles *7*. The hydrophobic light-scattering polyamide particles (PSP Dantex d=20 $\mu$m) of neutral buoyancy visualized the flow structure directly on the surface of the bubble. The addition of particles into the liquid volume visualized the flow around the gas inclusion. Before use, the aqueous solution of particles was examined for the presence of surfactant in a Langmuir–Blodgett trough and using the maximum bubble pressure method. The measurements showed that the addition of particles does not lead to contamination, which enables its use in research.

## 3. Results

Figure 3 presents the structure of the near-surface flow on the bubble and in the bubble, wake obtained by the dye-streak method at Reynolds numbers of 60 (figure 3(a)) and 120 (figure 3(b)). The bubble diameter and the characteristic velocity of the axisymmetric flow far from the bubble was used to calculate the Reynolds number. As one can see, at low Reynolds numbers (figure 3(a)), flow view as a creeping takes place in the system. All the jets on the bubble surface are well visualized. The width of the dye-streak varies, which allows us to estimate the flow velocity near the bubble surface relative to the surrounding flow. As the Reynolds number increases downstream, the formation of an attached vortex is observed behind the sphere (figure 3(b)). In the absence of a flow around the bubble (uniform dye supply), the formation of a vortex flow on the surface of the sphere was observed during the experiment, which can be attributed to the high density of the dye solution compared with the base liquid. Solution feeding generates additional disturbances in the system, which causes a vortex motion to occur on the surface of the gas inclusion and makes it impossible to use this visualization method in further research.

Usually, the spiral trajectory of the rising bubbles is explained by the loss of symmetry of the attached toroidal vortex. In our opinion, that the loss of mechanical equilibrium in the surfactant film and the appearance of the first instability mode in the form of two vortices on the bubble surface can lead to the same thing, yet at significantly lower Reynolds numbers. The experiments with rising bubbles in isopropanol and water have shown that the form of the rising trajectories changes from linear to spiral at equal Reynolds numbers; this can be directly associated with the mobility of the bubble interface. In the experiments with a stationary bubble in a water flow (at Re=30), in the wake flow behind the sphere an attached vortex is formed (view in figure 4(a)), while in the ispropanol flow it is absent (flow vision figure 4(b)). The presence of an attached vortex allows us to assess the mobility of the interface boundary. It is known from the literature that the attached vortex occurs in the flows around the solid sphere at Re>25. In the ideal case of an absolutely mobile bubble surface (the interface is free of surfactant molecules), the formation of an attached vortex should be observed at higher values.

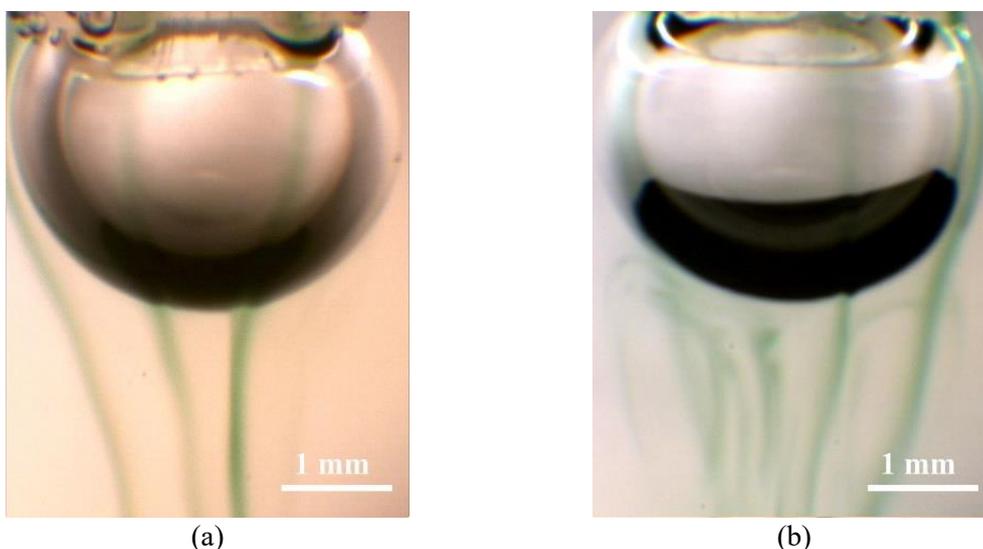

(a)          (b)

**Figure 3.** Liquid flow patterns obtained by ejection of dye Reynolds number: 60 (a) and 120 (b).

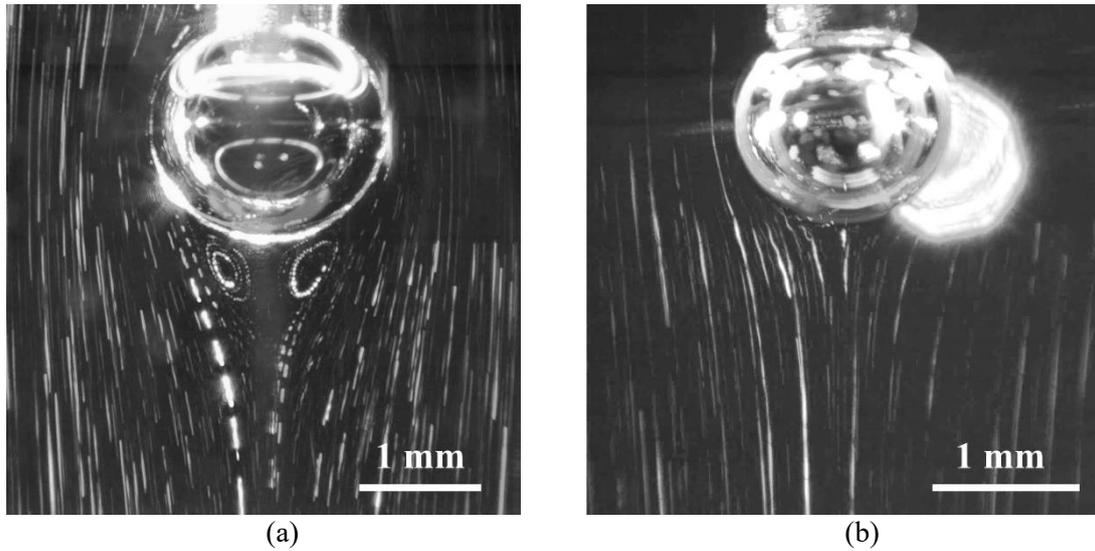

**Figure 4.** Plan view of the axisymmetric liquid flow around a gas bubble in water (a) and isopropanol (b) with Reynolds number 30.

The intensity of the thermocapillary flow near the bubble, formed in the absence of an axially symmetrical flow and on heating the end of capillary, is another sign of a clean surface (figure 5). The structures of the thermocapillary flow around the bubble realized in water and isopropanol are presented in figure 5(a) and figure 5(b), respectively. The application of a laser beam provides both visualizing the flow structure and local heating of the capillary. The temperature difference between the capillary and the fluid is less than one degree Celsius. Presented image is the tracks summing for 3-5 seconds after the bubble creating. In figure 5, the arrows indicate the direction of fluid motion. In both cases, a toroidal vortex is formed along the heated perimeter of the capillary. In the area under the bubble, an upward free-convective motion of the fluid` is observed. The track images clearly demonstrate the differences in the intensity and shape of the thermocapillary vortex and the thermogravitation flow under the gas sphere. The crossing tracks of the vortex near the equator of the bubble created in the water (figure 5(a)) indicate a change in the size of the vortex over time. This is caused by the adsorption of residual surfactant molecules from water to the interface; these molecules are transported by the thermocapillary flow to the lower pole of the bubble and, as a result, a stagnant zone is formed [3–5]. The vortex shape allows us to determine the position of the boundary of the moving part of the surface and to evaluate the significance of tangential stresses. It can be seen that in the case of isopropanol, the thermocapillary flow covers most of the surface and almost reaches the lower pole.

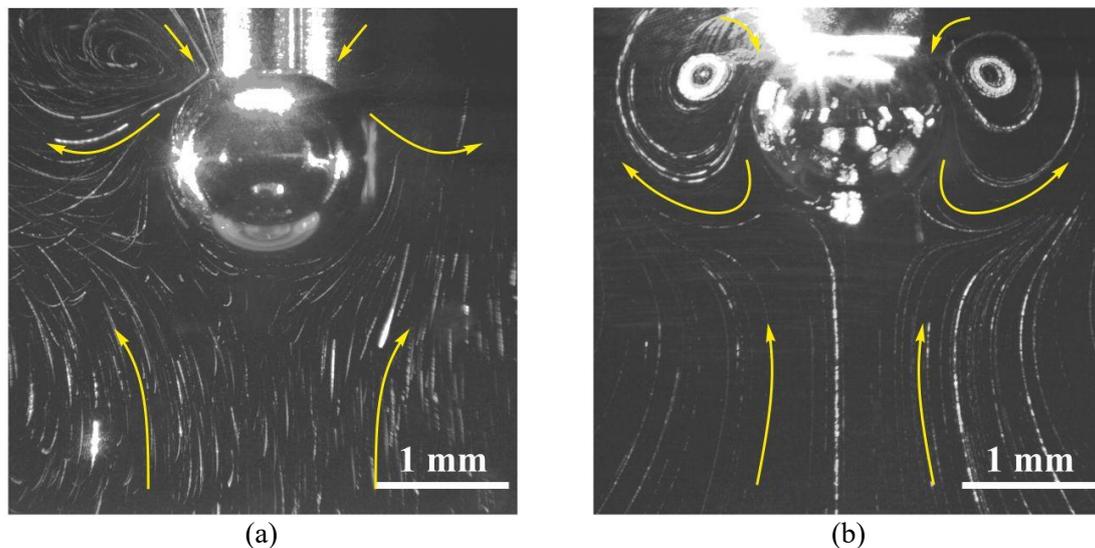

**Figure 5.** Structures of the thermocapillary flow around the bubble in water (a) and in isopropanol (b).

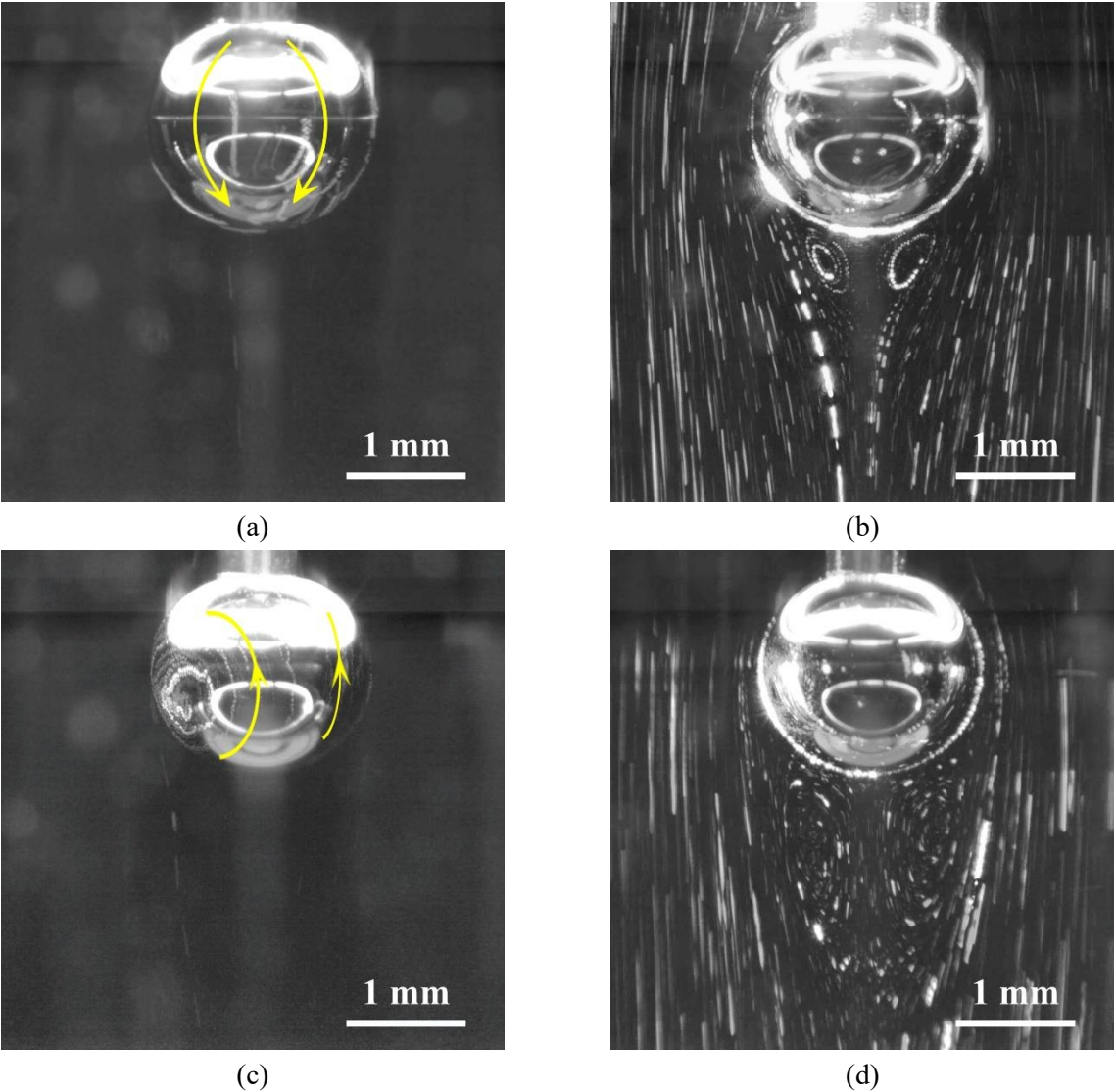

**Figure 6.** Flow structure on the bubble surface in diffuse light (a, c) and around it in laser beam (b, d) in water with Reynolds number 30 (a, b) and 80 (c, d).

The experimental results from figure 4 and figure 5 indicate that at this stage of the study, we have been able to realize the ideal case of a clean interface only using isopropanol, which is less sensitive to residual impurities. Despite all the techniques used to clean the system, the surface of the gas sphere in water experiments is contaminated, which causes its freezing.

The formation of an adsorption film of residual impurities on the surface of the gas inclusion in water is a sufficient condition for conducting experimental study of the flow on the bubble surface streamlined by uniform fluid flow. Figure 6, show the flow structure on the bubble surface in diffuse light (figure 6(a, c)) and around it (figure 6(b, d)) in laser beam for Reynolds numbers of 30 and 80, respectively. The images illustrate a typical scenario of instability development observed in various implementations when the Reynolds number reaches its critical value. At low Reynolds numbers, the tracers move at the bubble interface along the meridians, concentrating in the lower hemisphere (figure 6(a)). After that, they freeze and retain their immobility; the velocity of the axially symmetrical uniform flow remains unchanged. With increasing flow intensity, one can observe the formation of the first instability mode in the form of a two-vortex motion on the bubble surface (figure 6(c)). The axis of rotation is located near the equator and is orthogonal with respect to the velocity vector of the main flow. At the same time, the size of an attached vortex propagating downstream increases (figure 6(b) and 6(d)).

## 4. Conclusion

We have performed an experimental study to identify and investigate the conditions for the development of flow structure instability on the surface of the spherical gas inclusion interacting with the axially symmetric homogeneous flow. Our experiments were performed with a motionless gas inclusion in the laboratory coordinate system.

It is shown that the variation of the bubble size, the flow velocity, and the degree of contamination of the system have an impact on the surface flow structure. It has been found, at certain critical value of the control parameters, a two-vortex flow (the first mode of instability) is formed on the surface of a gas inclusion. The observed instability is consistent with the experimental results obtained for a flat surface [6, 7, 13]. Our research enables narrowing the range of variable parameters used in the numerical experiment and provides a deeper insight into the phenomena taking place at and nearby the interface between two phases. In our opinion, in the case of a free rising bubble of small diameter, the observed instability is responsible for the spiral rising trajectory. For bubbles with a diameter of 1-2 mm, this type of motion still has no generally accepted explanation, in contrast to larger bubbles, where the deviation from the spherical shape has a significant effect on the change of trajectory.

The results obtained demonstrate that the existing mathematical model for one of the basic problems of interface hydrodynamics about the rising of a single bubble needs to be refined. The presence of vortex structures on the surface of the rising bubble affects the interaction of several bubbles with each other and with a fine solid fraction in the flow, which is the basis of such technologies as flotation, extraction, and others related to interfacial mass exchange. In the future, this study can be useful to improve the efficiency of such technologies.

**Acknowledgement**
This research was supported by Russian Science Foundation (Grant No.: 19-71-00097).